# Rotating spintronic terahertz emitter optimized for microjoule pump-pulse energies and megahertz repetition rates

Alkisti Vaitsi,[1] Vivien Sleziona,[1] Luis E. Parra López,[1] Yannic Behovits,[2] Fabian Schulz,[1,3] Natalia Martín Sabanés,[1,4] Tobias Kampfrath,[2] Martin Wolf,[1] Tom S. Seifert,[2] and Melanie Müller[1*]

[1]Department of Physical Chemistry, Fritz Haber Institute of the Max Planck Society, 14195 Berlin, Germany
[2]Department of Physics, Freie Universität Berlin, 14195 Berlin, Germany
[3]CIC NanoGUNE, 20018 Donostia – San Sebastián, Spain
[4]IMDEA Nanoscience, 28049 Madrid, Spain

*Corresponding author: m.mueller@fhi-berlin.mpg.de

**Spintronic terahertz emitters (STEs) are powerful sources of ultra-broadband single-cycle terahertz (THz) field transients. They work with any pump wavelength, and their polarity and polarization direction are easily adjustable. However, at high pump powers and high repetition rates, STE operation is hampered by a significant increase in the local temperature. Here, we resolve this issue by rotating the STE at a few 100 Hz, thereby distributing the absorbed pump power over a larger area. Our approach permits stable STE operation at a fluence of ~1 mJ/cm$^2$ with up to 18 W pump power at megahertz repetition rates, corresponding to pump-pulse energies of a few 10 μJ and a power density far above the melting threshold of metallic films. The rotating STE is of interest for all ultra-broadband high-power THz applications requiring high repetition rates. As an example, we show that THz pulses with peak fields of 10 kV/cm can be coupled to a THz-lightwave-driven scanning tunneling microscope at 1 MHz repetition rate, demonstrating that the rotating STE can compete with standard THz sources such as LiNbO$_3$.**

Spintronic terahertz (THz) emitters (STE) are novel and versatile sources delivering ultra-broadband single-cycle THz pulses without spectral gaps [1] and with peak electric fields up to ~1 MV/cm [2,3]. Among these various advantages [4,5], STEs offer easy and versatile polarity and polarization control via external magnetic fields [6–8], wavelength-independent excitation [9] without phase matching constraints, excellent beam quality and focusability [2], and Fourier-limited single-cycle transients with ultra-wide bandwidth [1]. These features make STEs particularly interesting for THz field-driven applications such as THz-lightwave scanning tunnelling microscopy (THz-STM) [10–12] or field-resolved THz scanning near-field optical microscopy (THz-SNOM) [10,13]. A crucial aspect for these applications is the generation of measurable THz field-induced currents or scattered THz near-field signals, which requires operation at MHz repetition rates and sufficiently high THz field strength, e.g. up to few kV/cm in the case of THz-STM [10,14,15]. Modern Yb-based high-power femtosecond laser systems [16–18] combined with external pulse compressors [19–22] provide pulses with a duration down to a few 10 fs at several 10-100 W of output power and megahertz repetition rates, motivating the development and optimization of broadband THz sources such as the STE for such laser parameters [23–27]. This progress will pave the way towards future technological applications in broadband THz imaging [28,29], sensing and spectroscopy that will greatly benefit from high THz powers.

To exploit the full potential of the STE for field-driven THz applications, the conversion of optical pulse energy into emitted THz field (optical-to-THz field conversion efficiency, FCE) must be optimized as well as the propagation and focusing of the broadband THz pulses to the location of the experiment. Optimum FCE of the STE is typically achieved by operating the STE close to a pump fluence of ~1 mJ/cm$^2$ [2,23]. However, at pump powers of several 10 W and repetition rates in the megahertz range, accumulated heating over many laser pulses can significantly reduce the FCE and even lead to irreversible degradation or immediate thermal damage of the STE [23]. Excitation of the STE with pump spot sizes of several centimeters, as shown for multi-Watt operation at kilohertz repetition rates [2,3], minimizes heating and results in an optimal FCE at millijoule pump-pulse energies. However, at megahertz repetition rates and few 10 μJ pulse energies, this approach will result in very low fluence and correspondingly reduced FCE. On the other hand, if the fluence is increased by decreasing the pump spot size, the elevated power density and resulting significant heating will limit STE operation to low pump power. Thus, achieving a fluence of 1 mJ/cm$^2$ at megahertz repetition rates and multi-Watt average powers is a major challenge and requires efficient heat management. Active backside cooling of the STE was demonstrated recently [23], but requires operation in a back-reflection geometry that is less straightforward than a transmission geometry. In addition, the maximum achievable power will be limited by the heat transport within the cooled STE device, and scaling to higher powers will require complex cryogenic environments.

Here, we demonstrate efficient high-power operation of a trilayer STE excited at megahertz repetition rates with up to 18 W pump power and a pump spot size of a few millimeters, resulting in a fluence of ~1 mJ/cm$^2$ and pump power densities of ~350 W/cm$^2$, well above the damage threshold of thin metallic films. This progress is achieved by rotating the STE at an angular speed of 100 Hz, which effectively reduces the locally accumulated heat by distributing the absorbed pump power over an area much larger than the excitation spot. In addition to the THz generation process in the STE, achieving high THz electric fields at the point of the experiment requires an optimum THz-beam guiding. Accordingly, we optimize the collimation, propagation and focusing of the THz beam generated by the rotating STE. This procedure allows us to generate 10 kV/cm incident THz peak fields at the tip of a low-temperature STM, resulting in several Volts peak THz bias voltages between the STM tip and a metallic sample.

Figure 1 shows the experimental setup. We excite the STE with 1030 nm near-infrared (NIR) laser pulses of ~35 fs duration. The NIR pulses are generated by a high-power femtosecond laser (Light Conversion Carbide-80W), the output pulses (~200 fs) of which are spectrally broadened and temporally compressed in a multi-plate compressor (MPC) [19,20,22]. The system offers two operation modes, providing either 50 µJ or 20 µJ of compressed NIR pulse energy at a repetition rate of 1 MHz or 2 MHz, respectively. Part of the NIR power is split off for use as sampling pulses in electro-optic sampling (EOS) or for phase-resolved THz waveform sampling inside the STM [11,30]. A large-area STE (2" diameter, TeraSpinTec GmbH) is excited ~20 mm away from its center under normal incidence. The diameter $2w_p$ (1/e$^2$ of the intensity) of the collimated pump beam can be chosen to be 1.7, 2 and 3.7 mm by using different lenses in the beam path. The NIR pump power $P_p$ is controlled by a half-wave plate and a pair of thin-film polarizers. The STE is mounted on the shaft of a DC motor and can be rotated at a maximum speed of $f_{\rm rot} = 300$ Hz. It is magnetized by a large 25 mm diameter stationary permanent magnet mounted above it.

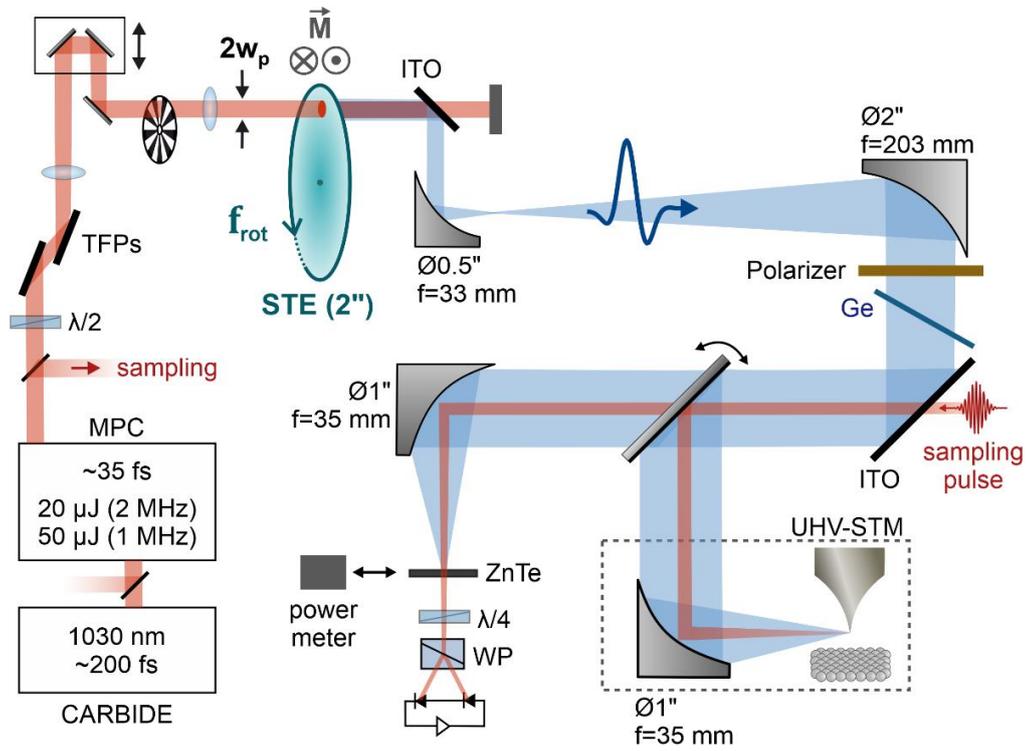

**Figure 1.** Experimental setup. $\vec{M}$: magnetization, MPC: multi-plate compressor, TFP: thin film polarizer, ITO: indium tin oxide glass, ZnTe: Zinc Telluride crystal for EOS, WP: Wollaston prism, Ge: Germanium wafer, UHV: ultrahigh vacuum, $w_p$: pump-beam radius ($1/e^2$ of intensity).

A glass plate coated with indium-tin-oxide (ITO) is mounted closely behind the STE to separate the THz radiation from the residual pump beam. The reflected THz pulses are collected by a 90° off-axis parabolic mirror (OAPM) which is part of a telescope (1:6) to expand the THz beam. A Germanium (Ge) wafer is installed at an angle of 55° to block the residual pump light. A second ITO-coated glass plate is used to overlap the NIR sampling pulses collinearly with the THz beam. A flip mirror allows us to send the THz beam either into the STM or to an identical beam path for EOS. In both cases, the THz beam is focused using a 90° OAPM with a diameter of 1" and a focal length of 35 mm. EOS traces are recorded using a 300 μm-thick ZnTe(110) crystal and are deconvoluted with the detector response function to extract the THz electric field at the detector position [31–33]. The THz power is measured by a pyroelectric detector (Gentec THZ9B-BL-DA, 30 nW detection limit) at the position of the EOS crystal, for which the NIR beam and, thus, the THz beam are chopped at 5

Hz with a 50% duty cycle to allow lock-in detection. All given NIR pump powers are chopped values. A second Ge wafer is mounted directly on the THz power meter to block NIR background radiation. A THz polarizer is used to block the linearly polarized THz pulses to measure the power of thermal radiation $P_{\text{th}}$ emitted by the STE. Finally, we measure the temperature of the STE by a heat camera (Model FLIR InfraCAM SD).

First, we characterize the thermal emission and the steady-state temperature $T_0$ of the STE during stationary and rotating operation. Figure 2(a) shows a heat camera image of the stationary STE excited with $2w_{\text{p}} = 3.7$ mm (size indicated by the pink spot) at a pump power $P_{\text{p}} = 6$ W. The STE heats up to $T_0 \approx 125$ °C in the center of the pump beam. Although the STE is not immediately damaged in this state, its efficiency decreases irreversibly within a few minutes, or several hours to days at lower powers in the range down to 1 W.

Figure 2(b) shows the spatial distribution of $T_0$ for the STE rotating at $f_{\text{rot}} = 300$ Hz. Remarkably, even at such comparatively low rotational frequencies, the heat is efficiently distributed over an annulus that is defined by the diameter and radial position of the pump beam and the thermal transport in the film. The temperature reaches $T_0 \approx 31$ °C at the position of the pump beam and decreases slightly in the direction of rotation. We further characterize the heating of the STE by blocking the linearly polarized THz pulses with the polarizer. In this case, the residual power reaching the power meter is dominated by the thermal radiation $P_{\text{th}}$ emitted by the heated STE. Figure 2(c) shows $P_{\text{th}}$ (right ordinate) and $T_0$ (left ordinate) measured over a large power range at $f_{\text{rep}} = 2$ MHz. For the stationary STE, $P_{\text{th}}$ increases steeply and similarly to $T_0$, where we limit the maximum power to about 6 W to avoid damaging the STE. When rotating the STE, the effectively heated area $A$ increases by a factor of $A_{\text{beam}}/A_{\text{ring}} \approx 10.5$ and the temperature decreases by a factor of $T_{0,\text{stat}}/T_{0,\text{rot}} \approx 4$ at 6 W pump power. Remarkably, $T_0$ stays below 50 °C up to about 16 W, and $P_{\text{th}}$ decreases considerably over the full power range. This clearly shows that even a moderately fast rotation of the STE can successfully reduce its average power heating.

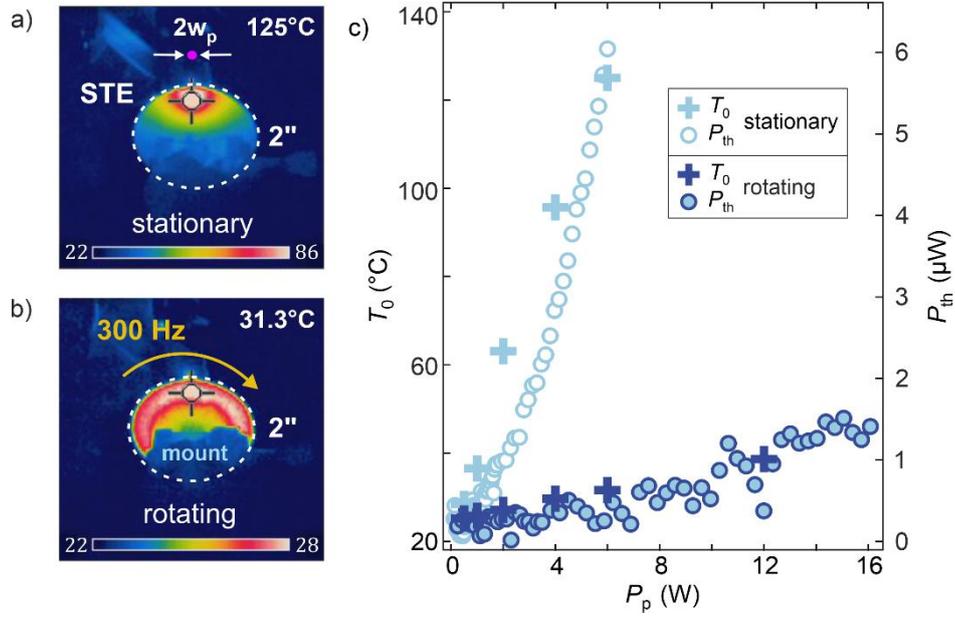

*Figure 2.* Spatial temperature profile of the (a) stationary and (b) rotating STE for $P_p = 6$ W. Note the rescaling of the color range. The pump spot has a $1/e^2$-intensity diameter of $2w_p = 3.7$ mm (size indicated by the pink dot in a)) and is located at a radial position of about 20 mm away from the STE center. (c) Steady-state STE temperature $T_0$ and total emitted power $P_{th}$ measured at the position of the EOS crystal for $f_{rep} = 2$ MHz and $f_{rot} = 300$ Hz.

Next, we analyze the dependence of the power $P_{\text{THz,STE}}$ of the coherently emitted THz pulses on the pump power $P_p$ for different pump-beam sizes and for the stationary and rotating STE [see Figure 3(a)]. Note that $P_{\text{THz,STE}}$ is here corrected for the optical transmission of the Ge wafers and the polarizer and the thermal radiation $P_{th}$ (see Supplementary Material). For the stationary STE (open markers), $P_{\text{THz,STE}}$ increases with increasing $P_p$ up to $P_p \approx 2$ W and then saturates or even decreases. This behavior is more pronounced for smaller $w_p$ as expected from the correspondingly higher power density. When the STE rotates, $P_{\text{THz,STE}}$ initially scales quadratically with $P_p$ up to a power that depends on $w_p$, and continues to increase linearly at higher powers. Moreover, $P_{\text{THz,STE}}$ is overall higher for smaller pump beams due to the higher fluence and correspondingly higher optical-to-THz power conversion efficiency (PCE) as expected for a second-order nonlinear-optical process. More information on the PCE is available in the Supplementary Material [23]. The results in Fig. 3(a) further show that the STE performance does not depend on the rotation frequency for $f_{rot}$ between 100 Hz and 300 Hz, indicating that lateral heat transport out of the excited region proceeds on ms time scales.

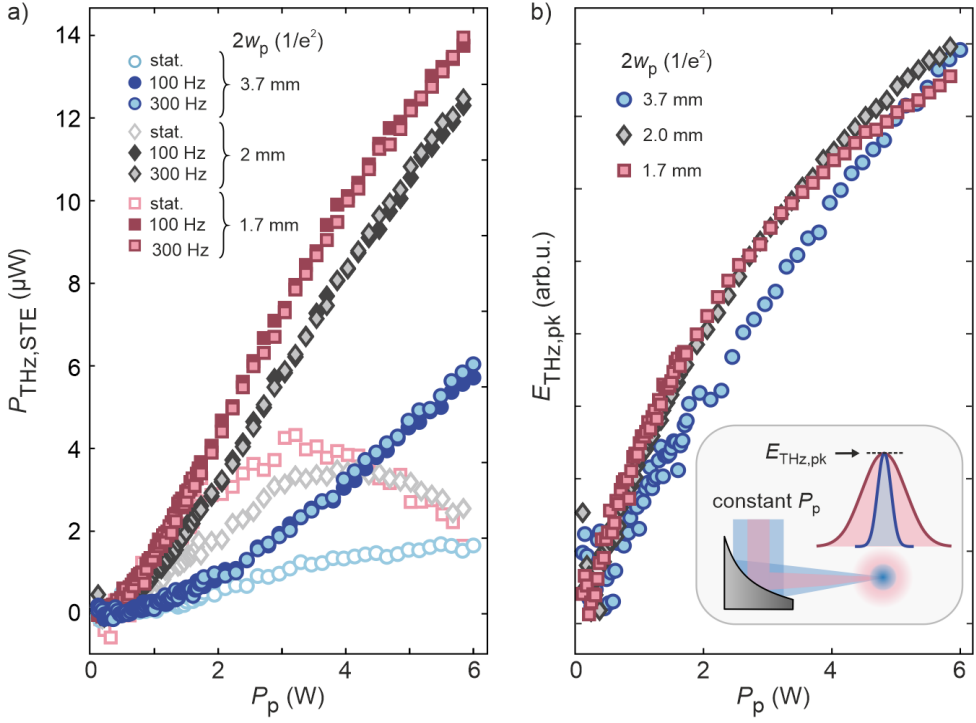

**Figure 3.** (a) THz power measured at the EOS position versus pump power $P_p$ for the stationary and rotating STE and for three pump spot sizes. (b) Peak THz electric field versus pump power for the same pump-spot sizes and $f_{rot} = 300$ Hz. Inset: illustration of constant THz peak field at constant pump power for varying pump spot size. In both panels, $f_{rep} = 2$ MHz.

For field-driven applications such as THz-STM, the incident THz field is a more relevant quantity than the THz power. Figure 3(b) shows the pump-power scaling of the THz peak field $E_{\text{THz,pk}}$ at the EOS crystal for the three pump-spot sizes and pump powers up to 6 W for $f_{\text{rot}} = 300$ Hz. In contrast to $P_{\text{THz,STE}}$, which at constant pump power increases with decreasing pump beam diameter due to the higher fluence as explained above [Fig. 3(a)], $E_{\text{THz,pk}}$ does not depend on $w_p$ and remains almost constant at a given pump power. This behavior is expected in the regime in which $E_{\text{THz}}$ scales linearly and, thus, $P_{\text{THz,STE}}$ quadratically with $P_p$, i.e., in the absence of heating and saturation effects. This effect can be understood by considering that the higher THz power emitted at small pump spot size (i.e., at higher fluence) is exactly compensated by the reduced THz beam diameter and the resulting larger THz focus at the EOS crystal, as sketched in the inset of Fig. 3(b). We support this explanation with a simple calculation in the Supplementary Material.

The fact that $E_{\text{THz,pk}}$ is approximately independent of $w_p$ further demonstrates that the THz pulses propagate through the setup without major losses due to apertures. The slightly smaller $E_{\text{THz,pk}}$ at $2w_\mathrm{p} = 3.7$ mm suggests that the THz beam is slightly clipped by the optical elements at this spot size, as further substantiated by the larger horizontal width of the THz focus [inset of Fig. 4(c)]. Finally, at higher powers above approximately 2 W, sub-linear scaling is observed for the smaller spot sizes, while for the largest spot size $E_\mathrm{THz,pk}$ scales nearly linearly with $P_\mathrm{p}$ in the investigated power range.

The data in Figure 3 shows that, for a given pump power, small spot sizes and higher fluence is advantageous for achieving higher THz powers. However, when high THz fields are required at low THz powers, as for example in THz-STM, larger pump spot sizes are advantageous as long as the THz beam propagates through the setup without clipping. The optimal condition for achieving the highest FCE and THz peak fields are obtained for pump spot sizes that ensure an excitation fluence of ~1 mJ/cm² and using a telescope with a magnification that allows the full available aperture of the setup to be used. However, a critical limit is reached when the beam size becomes comparable to the wavelength of the lower THz frequencies within the STE spectrum. In this case, low and high THz frequencies would propagate at drastically different divergence angles, making broadband collection and collimation of the pulses difficult (see Supplementary Material in [34]). Together with the single-pulse saturation limit imposed by the fluence [23], this behavior will limit the performance of the STE at smaller pump spot sizes.

Finally, we quantify the THz electric-field amplitudes we achieve in the experiment using the rotating STE for pump powers up to 18 W and at two different repetition rates of 1 MHz and 2 MHz and for $2w_\mathrm{p} = 3.7$ and 1.7 mm, respectively. We determine the THz peak field by [2,24,35]

$$E_\mathrm{THz,pk} = \sqrt{\frac{W_\mathrm{THz} Z_0/n}{(\pi/4\ln 2) f_x f_y \int dt\, |\hat{E}_\mathrm{THz}(t)|^2}}, \qquad (1)$$

where $Z_0$ is the vacuum impedance, $f_x$ and $f_y$ is the intensity full width at half maximum (FWHM) of the THz focus in $x$ and $y$ direction, respectively, $W_\mathrm{THz} = P_\mathrm{THz}/f_\mathrm{rep}$ is the THz pulse energy at the EOS position, $f_\mathrm{rep}$ is the laser repetition rate, and $\hat{E}_\mathrm{THz}(t)$ is the electric field of a single THz pulse with the peak normalized to one. Note that, unlike the $P_\mathrm{THz,STE}$ in Fig. 3(a), the $P_\mathrm{THz}$ here is the THz power at the EOS position and not at the STE, and $W_{\text{THz}}$ in Eq. (1) is the actual pulse energy incident on the EOS crystal or in the STM. We measure the THz spot size at the EOS position by the knife-edge method, and $f_x$ and $f_y$ are shown in the inset of Fig. 4(c). We confirm that the values for $E_{\text{THz,pk}}$ obtained by eq. (1) are in reasonable agreement with those obtained from calibration of the balanced detection signal in EOS (see Supplementary Material).

Figures 4(a) and 4(b) show the normalized THz electric field waveforms $E_{\text{THz}}(t)$ and their frequency spectra obtained by deconvoluting the EOS signal with the ZnTe detector response (see Supplementary Material). Figure 4(c) shows the scaling of the peak THz field $E_{\text{THz,pk}}$ (left ordinate) at the EOS position with pump-pulse energy $W_p$. At the highest available $W_p$ of about 36 µJ at 1 MHz, we achieve a THz peak field of ~10 kV/cm. Note that, at $2w_p = 1.7$ mm, we did not use the full available power to limit the maximum fluence and avoid laser pulse-induced optical damage [36].

Without saturation effects and aperture losses, we expect a constant $E_{\text{THz,pk}}$ for a constant pump-pulse energy independent of the pump fluence. Figure 4(c) confirms the expected behavior in the limit of small pulse energies. For low $W_p$ up to few microjoules, $E_{\text{THz,pk}}$ scales linearly with $W_p$ with a conversion factor that is independent of $f_{\text{rep}}$ and $2w_p$. At higher pulse energies, a sub-linear scaling is observed that is slightly more pronounced at 2 MHz, which we ascribe to residual pulse-to-pulse accumulation effects. At $f_{\text{rot}} = 300$ Hz and 2 MHz repetition rate, about 6600 pulses arrive at the STE during one revolution, and the STE rotates by an arc length of ≈17 µm between two consecutive pulses. Thus, ≈100 (≈215) pulses arrive in an excited area of 1.7 mm (3.7 mm) before a completely 'fresh' STE region has been moved into the pump spot. This condition appears to be sufficient to cause accumulation effects. Its understanding requires detailed knowledge of the heat transport within the STE, which is beyond the scope of this work. A higher rotational speed in the kilohertz range combined with a larger STE should allow for a complete elimination of accumulation and residual heating effects by operating the STE in the megahertz single-pulse regime.

Figure 4(c) (right ordinate) shows the THz peak bias voltage $U_{\text{THz}}$ that is induced between the STM tip and a metallic sample. The calibration of $U_{\text{THz}}$ is described in previous work and in the

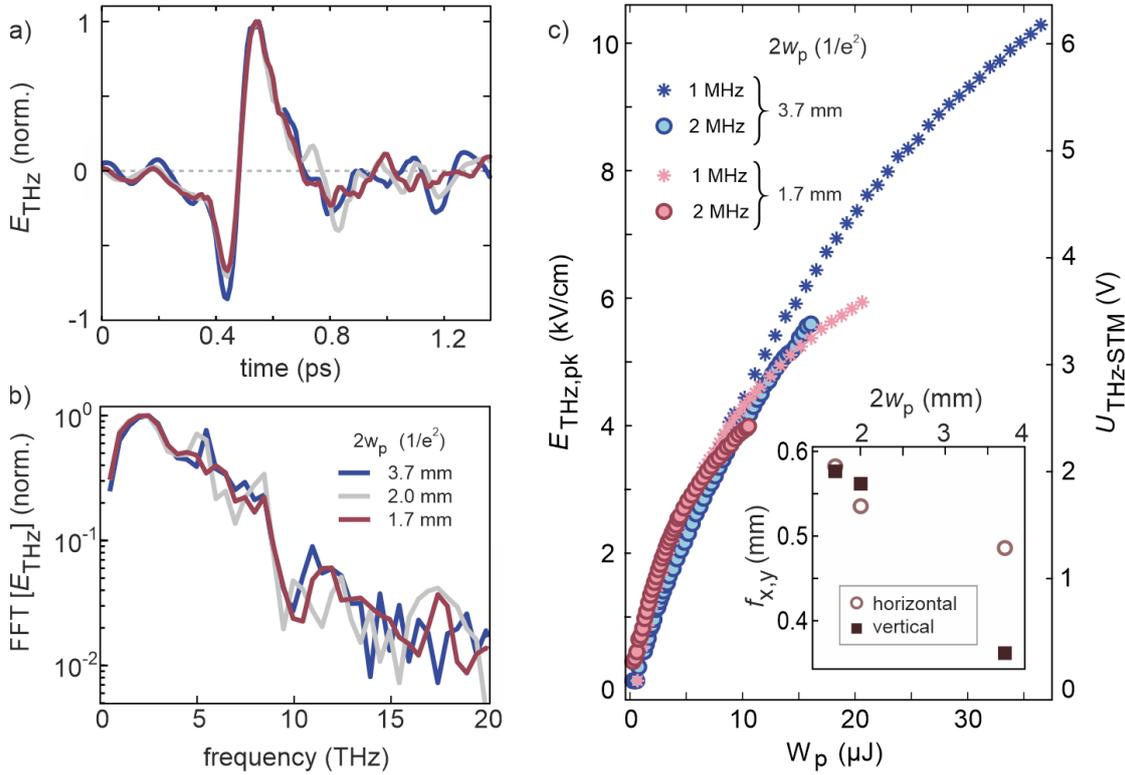

*Figure 4. Quantification of the THz electric field at the EOS position for two different repetition rates and pump-spot sizes. (a) Normalized THz electric-field waveforms and (b) corresponding THz spectra for $f_{rep}$ = 2 MHz. (c) Dependence of the THz peak electric field (left y-axis) and THz peak bias in the STM (right y-axis) on the pump pulse energy. Inset: Dependence of THz beam size ($f_x$, $f_y$) on pump-spot size. In all panels, $f_{rot}$ = 300 Hz.*

Supplementary Material [11,30]. The rotating STE allows us to generate peak THz bias voltages up to ≈6.2 V for the tip used in this work. Under optimal alignment conditions, the conversion of $E_{THz}$ into local THz bias depends on the geometry and material of the STM tip and sample. For a standard tungsten tip over a metal surface, we find a conversion factor of ~0.6 V/(kV/cm) similar to previously reported values [14]. We emphasize that we achieve several Volts of THz bias in the STM at THz pulse energies of only few 10 pJ, much lower compared to similar setups using LiNbO$_3$. This feature is advantageous for heat-sensitive field-driven applications such as THz-STM and demonstrates the high quality of the THz focus and THz beam emitted by the STE.

In conclusion, we demonstrate a rotating spintronic THz emitter that can be operated at high pump-power densities far above the damage threshold of thin metal films. Such rotating STE operation is enabled by the large-area homogeneity of the thin-film growth used for STE fabrication, allowing stable THz emission with no measurable fluctuations during rotation. We find that rotating a large-area STE at a few 100 Hz is sufficient to efficiently distribute the power over a large annulus area, thereby reducing the locally absorbed power density. This approach allows us to excite the STE by laser pulses of several 10 µJ energy within a collimated beam of few millimeters diameter at megahertz repetition rates, facilitating STE operation at an optimal fluence of ~1 mJ/cm$^2$ using high-power, high-repetition rate laser systems. The rotating STE is of particular interest for field-driven THz applications that require high repetition rates and reasonably high THz fields, for example, THz-STM. Finally, our results demonstrate that the rotating STE can easily compete with standard single-cycle THz sources such as LiNbO$_3$ for such applications. Furthermore, the design is scalable, and using a larger STE and faster rotational speeds should allow STE operation at much higher pump powers up to the kilowatt-level.

## SUPPLEMENTARY MATERIAL

Supplementary materials available: 1) THz power calibration; 2) Thermal radiation; 3) Power conversion efficiency at STE; 4) Electro-optic sampling; 5) Dependence of the THz electric field on pump spot size; 6) THz bias calibration in STM.

## ACKNOWLEDGEMENTS

All authors thank O. Gückstock and S. Mährlein for fruitful discussion, and H. Oertel and D. Wegkamp for technical support. M.M., A.V., V.S., and L.E.P.L., thank the Max Planck Society for financial support. T.S.S. and T.K. acknowledge funding by the Deutsche Forschungsgemeinschaft (DFG, German Research Foundation) through the Collaborative Research Center SFB TRR 227 "Ultrafast spin dynamics" (Project No. 328545488; projects A05, B02 and B05) and through the priority program SPP 2314 "INTEREST" (Project ITISA; Project No. KA 3305/5-1).

Supplementary Material

# Rotating spintronic terahertz emitter optimized for microjoule pump-pulse energies and megahertz repetition rates

Alkisti Vaitsi,[1] Vivien Sleziona,[1] Luis E. Parra López,[1] Yannic Behovits,[2] Fabian Schulz,[1,3] Natalia Martín Sabanés,[1,4] Tobias Kampfrath,[2] Martin Wolf,[1] Tom S. Seifert,[2] and Melanie Müller[1]

[1]*Department of Physical Chemistry, Fritz Haber Institute of the Max Planck Society, 14195 Berlin, Germany*

[2]*Department of Physics, Freie Universität Berlin, 14195 Berlin, Germany*

[3]*CIC NanoGUNE, 20018 Donostia – San Sebastián, Spain*

[4]*IMDEA Nanoscience, Ciudad Universitaria de Cantoblanco, 28049 Madrid, Spain*

**Table of Contents**

1. **THz power calibration**

2. **Thermal radiation**

3. **Power conversion efficiency at STE**

4. **Electro-optic sampling**

5. **Dependence of the THz electric field on pump spot size**

6. **THz bias calibration in STM**

**1. THz power calibration**

The THz powers are measured with the Gentec THZ9B-BL-DA pyroelectric detector in lock-in configuration. To retrieve the THz power, we apply a set of calibration factors supplied with the detector: the device's voltage responsivity (in units V/W) and a multiplier to correct for the detector's wavelength-specific response at the peak of our THz spectrum (wavelength ~150 µm). Furthermore, all power values are corrected for the transmission of the Ge wafer (60% at normal incidence), which is mounted on the power-meter head, and for the transmission of the THz-polarizer (89% transmittance at a wavelength of ~150 µm), as these components are not part of the setup under normal experimental conditions but were only added for the THz power measurement. Finally, we subtract the thermal radiation (see next paragraph) from the measured total THz power. To specify the THz power directly behind the STE ($P_{\text{THz,STE}}$ in the manuscript) and to calculate its power conversion efficiency [Fig. 3(a) and Fig. S4], we correct for the losses at the second Ge wafer (85% transmittance for p-polarized light at the used angle of incidence) that is used in our setup to block residual pump light (see Fig. 1).

**2. Thermal radiation**

Figure S1 shows the thermal power $P_{th}$ emitted by the rotating STE that is incident on the detector position over the full range of available pump powers for 1 MHz and 2 MHz repetition rate and two pump spot sizes, respectively. The thermal power is the power measured by the THz power meter when the polarizer is oriented perpendicular to the linearly polarized THz pulses, At low pump powers (1 to 2 W), when there is negligible heating of the STE, no THz signal is detectable when the polarizer is oriented perpendicular to the linearly polarized THz pulses, verifying that the transmitted THz power in this case originates predominantly from thermal radiation. To specify the thermal power that is present in later experiments, when the polarizer is not installed, we correct the displayed thermal power values for the 50% loss of thermal photons at the THz polarizer.

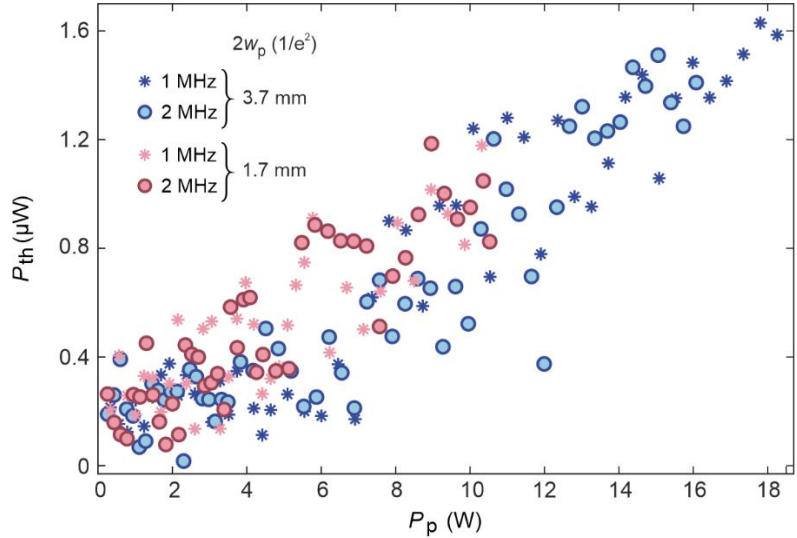

*Figure S1.* Thermal radiation power incident on the detector position versus average pump power for two repetition rates and pump spot sizes. ($f_{rot} = 300$ Hz).

The results in Figure S1 show that the thermal power does not depend on the laser repetition rate. The slightly higher $P_{th}$ found for the smaller pump spot size could originate from the smaller annulus area formed by the smaller pump spot, and hence a correspondingly slightly higher power density and temperature within the annulus where the STE is effectively excited.

## 3. Power conversion efficiency at STE

Figures S2(a) and S2(b) show the fluence scaling of the STE power conversion efficiency (PCE) and the THz peak field at the EOS position for 1 and 2 MHz repetition rate and for the two spot sizes of 3.7 and 1.7 mm, respectively. In the absence of saturation effects, a constant PCE is expected for constant fluence, which we observe in the limit of low fluences. At higher fluence, the PCE saturates at a value of $\approx 6 \times 10^{-6}$ at 1 MHz for both spot sizes, whereas a smaller maximum PCE is observed at 2 MHz (PCE saturation was only achievable for the high fluences at $2w_p =1.7$ mm), indicating that residual pulse-to-pulse accumulation effects slightly affect the PCE of the rotating STE at our conditions. Figure S2(b) shows that sub-linear scaling becomes

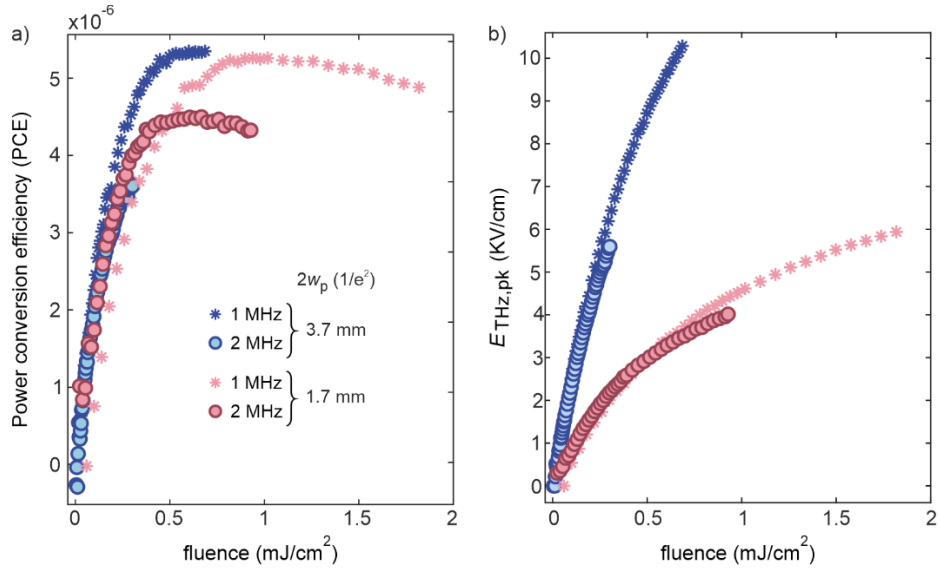

***Figure S2.*** *(a) Power conversion efficiency of the rotating STE versus NIR pump fluence for two repetition rates and pump spot sizes. (b) Fluence dependence of the peak THz field at the EOS position for the same experimental parameters. ($f_{rot}$ = 300 Hz)*

significant at fluences exceeding ≈ 0.2-0.3 mJ/cm². Note that higher THz fields are obtained at constant fluence for larger pump spot size due to the tighter THz focusing and hence smaller THz beam waist. Finally, Fig. S2(b) shows that the fluence scaling of $E_{\text{THz,pk}}$ is independent of the repetition rate.

## 4. Electro-optic sampling

Figure S3 shows the waveform (left) and spectrum (right) of the electro-optic sampling signal for the three spot sizes and data shown in Fig. 4 in the main manuscript.

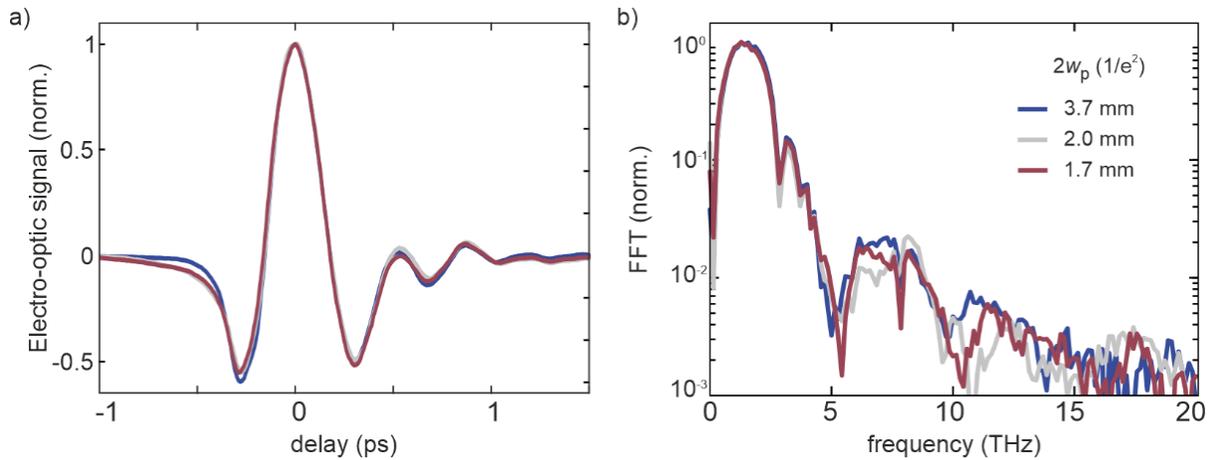

***Figure S3.*** *(a) Electro-optic signals, from which the THz electric field waveforms in Fig. 4a) are obtained by deconvolution with the detector response. (b) Corresponding THz spectra. ($f_{rep}$ = 2 MHz, $f_{rot}$ = 300 Hz)*

To cross-check the calculation of the THz field using Eq. (1) in the main text, we compare the values in Fig. 4(c) with the peak THz field values obtained from the calibration of the EOS signal measured via balanced detection. The EOS calibration factor is given by the ratio of the difference signal $\Delta J = J_1 - J_2$ and the sum of the two photodiode signals $J_1$ and $J_2$, i.e., by $\Delta J/(J_1 + J_2)$. We find slightly higher field values from calibration of the EOS measurements compared to those obtained from Eq. (1). For example, at 3 W pump power, 2 MHz repetition rate, and 3.7 mm pump spot, we obtain a peak THz electric field of 2.4 kV/cm from the EOS calibration and the balanced detection, compared to 1.4 kV/cm from the calculation using Eq. (1). Considering the uncertainties and inaccuracies in the measurements, these values are in good agreement. Even though all alignments and procedures have been carefully checked for robustness and accuracy, remaining uncertainties may arise, for example, from (i) the frequency-dependent response of the power meter (being more sensitive to higher frequencies), (ii) the resulting error in the determination of the THz focus by the knife-edge method (frequency-dependent focal size), (iii) the precise positioning of the ZnTe crystal or the knife-edge in the THz focus, or (iv) uncertainties in the determination of the THz-electric-field waveform.

## 5. Dependence of the THz electric field on pump spot size

As explained in the main manuscript and shown in Fig. 3(b), the peak THz field $E_{\text{THz,pk}}$ at the EOS position does not depend on the pump-spot size. This observation can be understood by a simple calculation assuming a pump beam with a simple circular flat top profile. We assume the simplest geometry in which the STE is excited by a collimated pump beam and the THz is focused by a single mirror in vacuum. If $r$ is the radius of the pump beam, we consider two cases: $r_1$ and $r_2$ with $r_1 = ar_2$ where $a$ is a constant coefficient. The waist of the focused THz beam is $w_1$ and $w_2$, respectively. Considering excitation of the STE at constant pump power $P_\text{p}$, the pump intensities for the two cases $r_1$ and $r_2$ are

$$I_{\text{p1}} = \frac{P_\text{p}}{\pi r_1^2} \quad \text{and} \quad I_{\text{p2}} = \frac{P_\text{p}}{\pi r_2^2} = a^2 I_{\text{p1}}.$$

The emitted THz fields scale linear with the pump fluence and intensity, so that

$$E_{\text{THz,1}} = bI_{\text{p1}} \quad \text{and} \quad E_{\text{THz,2}} = bI_{\text{p2}} = ba^2 I_{\text{p1}}$$

with the proportionality factor $b$. The corresponding THz intensities at the STE are

$$I_{\text{THz},1} = \frac{c\varepsilon_0}{2}|E_{\text{THz},1}|^2 = b^2 \frac{c\varepsilon_0}{2} I_{p1}^2 \quad \text{and}$$

$$I_{\text{THz},2} = \frac{c\varepsilon_0}{2}|E_{\text{THz},2}|^2 = b^2 \frac{c\varepsilon_0}{2} I_{p2}^2 = b^2 \frac{c\varepsilon_0}{2} a^4 I_{p1}^2$$

and the respective THz powers are

$$P_{\text{THz},1} = I_{\text{THz},1} \pi r_1^2 = b^2 \frac{c\varepsilon_0}{2} I_{p1}^2 \pi r_1^2 \quad \text{and}$$

$$P_{\text{THz},2} = I_{\text{THz},2} \pi r_2^2 = b^2 \frac{c\varepsilon_0}{2} a^4 I_{p1}^2 \pi r_2^2 = b^2 \frac{c\varepsilon_0}{2} a^2 I_{p1}^2 \pi r_1^2 = a^2 P_{\text{THz},1}.$$

In the paraxial approximation, the waist of a Gaussian beam is inversely proportional to the beam diameter entering the focusing lens. Hence, assuming the diameter of the THz beam is the same as that of the collimated pump beam, the waists of the two focused THz beams relate to each other as $w_2 = a w_1$, i.e., in an inverse manner compared to the pump radius $r_2 = a^{-1} r_1$. In other words, the larger the initial beam diameter, the smaller is its diameter after focusing. The peak THz fields at the center of the beam waist are then

$$E_{\text{THz},1} = \frac{2}{c\varepsilon_0} \frac{P_{\text{THz},1}}{\pi w_1^2} \quad \text{and}$$

$$E_{\text{THz},2} = \frac{2}{c\varepsilon_0} \frac{P_{\text{THz},2}}{\pi w_2^2} = \frac{2}{c\varepsilon_0} \frac{a^2 P_{\text{THz},1}}{a^2 \pi w_1^2} = E_{\text{THz},1}$$

In conclusion, as observed in the experiment, the THz peak fields in the focus are the same for both initial pump spot sizes. The higher THz power obtained for a smaller pump spot is exactly compensated by the smaller THz beam waist due to less tight focusing.

## 6. THz bias calibration in STM

The THz bias is calibrated as described in previous literature [1]. Figure S4(a) shows the photoemission current that is generated from the apex of a tungsten STM tip by the NIR sampling pulse versus the STM bias. Operating at high bias of 6 V, where the photoemission originates solely from the tip, the recorded THz waveform reflects the THz-induced change of the photoemission current. Due to the quasi-static nature of the THz field, we can calibrate the THz bias amplitude by dividing the measured THz-induced change of the photocurrent by the linear $I$-$V$ slope $a = 0.25$ pA/V of the DC photocurrent-voltage curve.

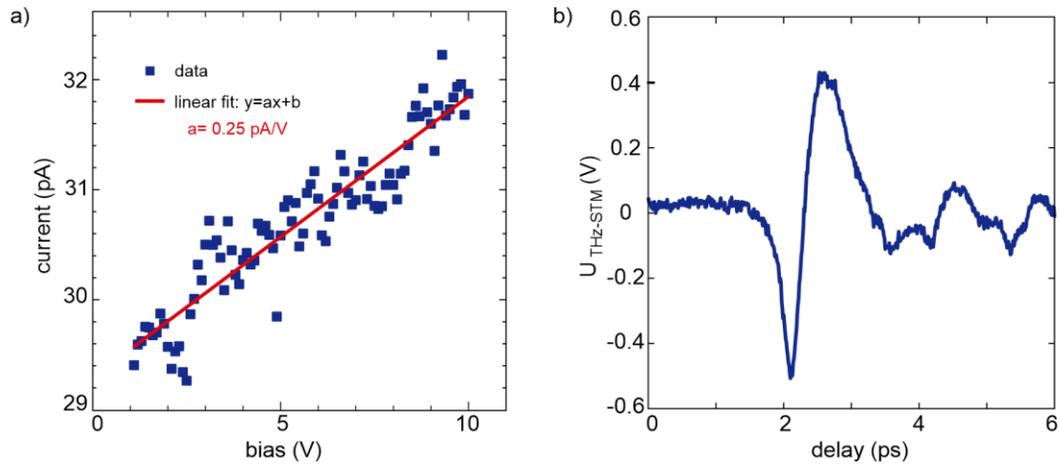

**Figure S.4** (a) Photocurrent-voltage curve of the photoemission current excited from the STM tip and used to calibrate the THz bias in Fig.4 (a) and (b) in the manuscript. (b) Calibrated THz bias transient measured at 6 V DC bias inside the STM. ($f_{rep} = 2$ MHz, $P_p = 2$ W)